\documentclass[preprint,12pt]{elsarticle}

\usepackage{hyperref}
\biboptions{sort&compress}

\usepackage{amssymb}
\usepackage{amsmath}
\usepackage{graphicx}

\usepackage{lineno}
\usepackage{xcolor}

\usepackage{siunitx}
\usepackage{subcaption}

\usepackage{array}
\usepackage{booktabs}

\newcolumntype{L}[1]{>{\raggedright\arraybackslash}p{#1}}

\journal{Nuclear Inst. and Methods in Physics Research, A}

\begin{document}

\begin{frontmatter}



\title{A muonium laser ionization system for muon acceleration studies at J-PARC}


\author[KEK]{Shusei Kamioka\corref{cor1}} 
\cortext[cor1]{Corresponding author}
\ead{kamioka@post.kek.jp}
\author[U-Tokyo]{Meng Lyu} 
\author[Ibaraki]{Katsuyoshi Homma} 
\author[KEK,U-Tokyo]{Tsutomu Mibe} 
\author[Ibaraki]{Shinki Nakamura} 
\author[KEK]{Yu Oishi\corref{cor1}}
\ead{bigstone@post.kek.jp}
\author[KEK]{Koichiro Shimomura} 
\author[Riken]{Tatsuya Shinozaki}
\author[KEK]{Hikaru Sunagawa} 
\author[Riken]{Satoshi Wada} 
\author[Ibaraki]{Takumi Yamada} 
\author[KEK]{Takayuki Yamazaki}

\affiliation[KEK]{organization={High Energy Accelerator Research Organization (KEK)},
            addressline={1-1 Oho}, 
            city={Tsukuba},
            postcode={305-0801}, 
            state={Ibaraki},
            country={Japan}}

\affiliation[U-Tokyo]{organization={Graduate School of Science, University of Tokyo},
            addressline={7-3-1}, 
            city={Hongo, Bunkyo-ku},
            postcode={113-0033}, 
            state={Tokyo},
            country={Japan}}

\affiliation[Riken]{organization={Center for Advanced Photonics, RIKEN},
            addressline={2-1}, 
            city={Hirosawa, Wako},
            postcode={351-0198}, 
            state={Saitama},
            country={Japan}}

\affiliation[Ibaraki]{organization={College of Engineering, Ibaraki University},
            addressline={4-12-1}, 
            city={Nakanarusawa, Hitachi},
            postcode={316-8511}, 
            state={Ibaraki},
            country={Japan}}

\begin{abstract}
This paper reports the development and commissioning of a muonium ionization laser system installed in the H2 area of the J-PARC Materials and Life Science Experimental Facility for muon accelerator studies. 
In this accelerator scheme, ultraslow muons generated by laser ionization of thermal muonium atoms serve as the particle source for subsequent radio-frequency acceleration.
The laser system generates coherent 122-nm pulses for muonium excitation and 355-nm pulses for photoionization of the excited muonium. 
A vacuum-ultraviolet (VUV) pulse transport system and VUV pulse diagnostics were also developed.
Muonium ionization was confirmed using a 5.7~keV muon transport line, and a muon rate of approximately 6~s$^{-1}$ was observed.
These results establish operation of the ionization system and provide a platform for subsequent muon accelerator studies.

\end{abstract}




\end{frontmatter}



\section{Introduction}
Low-emittance positive muon beams produced by accelerating ultraslow
muons are expected to enable a wide range of future experiments in
particle physics and materials science~\cite{10.1093/ptep/ptz030,TmuM,Shimomura:2025_muon_imaging}. 
In this approach, surface muons are stopped in a muonium-production target~\cite{PhysRevLett.74.4811}.
A fraction of the stopped muons forms thermal muonium atoms (Mu, $\mu^+e^-$), which are emitted into vacuum.
The emitted muonium atoms are then resonantly excited and photoionized by laser pulses, producing thermal positive muons.
These low-energy muons are referred to as ultraslow muons and are subsequently extracted and accelerated.
As a thermal particle source, ultraslow muons have a small normalized transverse rms emittance of approximately $0.3~\pi$mm$\cdot$mrad~\cite{10.1093/ptep/ptz030}, suitable for subsequent acceleration.
One of the proposed applications is the measurement of the muon anomalous magnetic moment and the search for the muon electric dipole moment~\cite{10.1093/ptep/ptz030}.
Another possible application is muon imaging using an accelerated muon beam.
A high-intensity muon beam with a well-defined energy would enable muon imaging with higher resolution and shorter acquisition times than imaging based on cosmic-ray muons~\cite{Otani:2021_muon_imaging,Shimomura:2025_muon_imaging}.
Recently, radio-frequency acceleration of ultraslow muons to 100~keV was demonstrated~\cite{PhysRevLett.134.245001}.

Following this demonstration, a new experimental area, called the H2 area, is being developed toward the implementation of a muon accelerator at the Materials and Life Science Experimental Facility (MLF) of the Japan Proton Accelerator Research Complex (J-PARC)~\cite{doi:10.7566/JPSCP.45.011045,Kamioka2026MuonAcceleration}.
A high-intensity surface muon beam from the H-line is transported to the area~\cite{10.1093/ptep/pty116}.
In the area, studies of 340~keV muon acceleration using a radio-frequency quadrupole (RFQ) and 4.3~MeV acceleration using an interdigital H-mode drift-tube linac (IH-DTL) are planned~\cite{PhysRevSTAB.16.040102, PhysRevAccelBeams.19.040101, PhysRevAccelBeams.25.110101,1l3t-rvmf}.
Recently, a low-energy ultraslow-muon transport and diagnostic line for the H2 area has been developed and commissioned~\cite{wada2026diagnostic57kevmuon}.

\begin{figure*}[tb]
\centering
\includegraphics[width=0.95\linewidth]{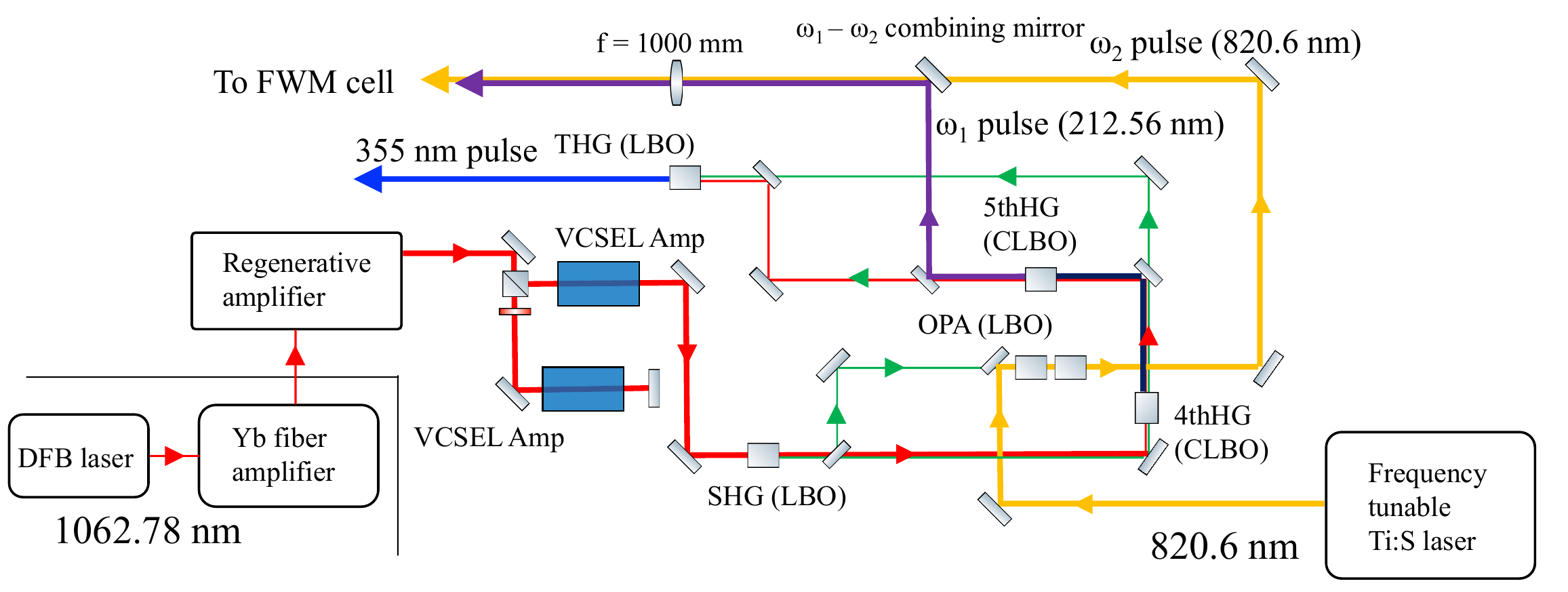}
\caption{
Schematic view of the 122-nm and 355-nm laser system.
The 212.56~nm \(\omega_1\) pulse is generated by amplifying 1062.78~nm light from a distributed-feedback semiconductor (DFB) laser and subsequently converting its frequency through harmonic generation. 
The 820.6~nm \(\omega_2\) pulse is generated by a frequency-tunable Ti:sapphire laser.
The $\omega_1$ pulse and the $\omega_2$ pulse are combined and focused into the FWM cell to generate the 122~nm pulse.
The 355~nm pulse is produced from residual pulses in the $\omega_1$ generation system.
SHG, THG, 4thHG, 5thHG, and OPA denote second-, third-, fourth-, and fifth-harmonic generation and optical parametric amplification, respectively.
VCSEL Amp denotes a vertical-cavity surface-emitting laser (VCSEL)-pumped amplifier.
}
\label{fig:light_source}
\end{figure*}

Laser ionization of muonium is one of the key components of the muon accelerator.
A resonant multiphoton ionization scheme is required to photoionize muonium, since its binding energy of 13.6~eV is so high that photoionization using a single photon is technically challenging.
In the previous muon acceleration experiment, a 244~nm light source originally developed for precision laser spectroscopy of muonium was used for muonium ionization via the 2$S$ state~\cite{PhysRevLett.134.245001,Kimura:2026_usm_property_s2}. 
The two-photon \(1S\)-\(2S\) transition requires a high laser intensity~\cite{Zhang:2021cba}, making it challenging to maintain sufficient excitation efficiency over the large transverse extent of the muonium distribution~\cite{Kamioka:2026_usm_simulation}.
In the previous acceleration experiment, the muon beam rate was $2\times 10^{-3}$~$\mu^+$/pulse~\cite{PhysRevLett.134.245001}. 
In addition, the 244~nm light source had a pulse duration of 60~ns, whereas a shorter pulse is required to reduce the time width of the accelerated muon beam~\cite{10.1093/ptep/ptz030}.

A light source capable of providing a higher muon rate and suitable for integration into the H2 muon accelerator apparatus is therefore required.
At the MLF U-line for $\mu$SR experiments with ultraslow muons, ultraslow muons are produced by laser ionization via excitation to the 2$P$ state~\cite{Kanda_2023}.
The \(1S\)–\(2P\) transition has a much larger transition dipole moment than the two-photon \(1S\)–\(2S\) transition, making this ionization scheme advantageous for efficient muonium ionization.
The same ionization scheme was therefore adopted for the H2 muon accelerator system.
However, the short-wavelength light source required for excitation to the 2$P$ state is technically challenging~\cite{BAKULE2008335,Saito:16}.
In addition, since 122~nm is in the vacuum-ultraviolet (VUV) region, a dedicated VUV transport system under vacuum is also required.

In this paper, we report the development and commissioning of a muonium ionization laser system
for the muon accelerator installed in the H2 area of J-PARC MLF.
The laser system produces coherent 122~nm pulses for excitation of ground-state
muonium to the 2$P$ state.
A 355~nm pulse is also produced to ionize the excited muonium.
The basic light-source scheme was established by Saito et al.~\cite{Saito:16} and was adopted here as the basis of a new system developed for integration into the muon accelerator apparatus.
A VUV pulse transport system and its diagnostics have also been developed.
These diagnostics also provide laser parameters that contribute to characterizing the initial ultraslow-muon beam distribution for subsequent acceleration.

This paper is organized as follows. Section~\ref{chap:lightsource} describes the configuration of the light source developed for the H2 area. 
Section~\ref{chap:diagnostics} describes the VUV pulse transport system and VUV diagnostics. The characteristics of the VUV pulses are also discussed.
The results of the commissioning of the light source during beam time to demonstrate muonium ionization are reported in Section~\ref{chap:beamtime}.
Section~\ref{chap:discussion} discusses the commissioning results and the future upgrade plan, followed by the conclusion.

\section{Laser system}
\label{chap:lightsource}
A cleanroom has been constructed adjacent to the experimental area, next to
the radiation shield of the H2 area.
The laser system is installed on a 7.2~m $\times$ 0.8~m optical bench placed
along the wall of the cleanroom facing the radiation shield.
The optical axis is 1.6~m above the floor, matching the height of the beamline
exit.
The laser pulses are transported to the experimental area through an optical
penetration in the cleanroom and radiation-shielding walls.

Figure~\ref{fig:light_source} shows a schematic view of the laser system.
The laser system was newly constructed for the H2 area, adopting the basic light-source scheme established by Saito et al.~\cite{Saito:16,Oishi2014Amplifier} and further developed by Oishi et al.~\cite{Oishi_2023} for operation at the MLF U-line.
Compared with the previous system, several components of the light source and the VUV transport system were modified or newly designed for operation in the H2 area, as described below.

The 122~nm pulse is produced using a two-photon-resonant four-wave mixing (FWM) process in Kr gas~\cite{Marangos:90,PhysRevA.84.033842,10.1063/1.4963303}.
For FWM, two input pulses are required: two-photon-resonant excitation light at $\omega_1=212.56$~nm and mixing light at $\omega_2=820.6$~nm.
The generated VUV frequency is given by $\omega_{\mathrm{VUV}}=2\omega_1-\omega_2$.

The $\omega_1$ pulse is generated from a 1062.78~nm laser through harmonic conversion. 
The \(\omega_2\) pulse is generated from a frequency-tunable Ti:sapphire laser and amplified using an optical parametric amplifier (OPA). 
The 355~nm pulse is also generated from the 1062.78~nm laser system.
They are described in more detail in the following subsections.

\subsection{Generation of the $\omega_1$ pulse}
The $\omega_1$ pulse is produced as the fifth harmonic of pulsed light at 1062.78~nm.
A fiber-coupled distributed-feedback semiconductor (DFB) laser (QLD1061, QDLaser) is used as the master oscillator.
Its wavelength is monitored using a wavelength meter (WS6, HighFinesse) and tuned to 1062.78~nm.
The seed laser is amplified and pulsed by Yb fiber amplifiers developed by Shinozaki et al.~\cite{Shinozaki:25}.
The amplifiers are operated at a repetition rate of 100~kHz.
The trigger for the pulse generation is derived from the 12~MHz reference signal used for the accelerator RF system, so that the laser pulses are synchronized with the accelerator RF.
The pulse duration is controlled by the width of the trigger pulse,
which was set to 3.2~ns in this study.
The temporal pulse duration is measured to be 2.9~ns (FWHM) using a fast photodiode (FPD-VIS-300, Ophir) with a specified rise time of less than 0.3~ns and a 2-GHz-bandwidth oscilloscope.

The 1062.78~nm pulse is amplified using a regenerative amplifier. 
Nd:YGAG crystals are used as the amplifier gain medium ($4\times4\times10~\mathrm{mm}^3$)~\cite{Oishi2014Amplifier,Walsh:98}.
The amplifier is operated at 25~Hz, matching the repetition rate of the surface muon beam, and its output pulse energy is 6~mJ.
The pulse is further amplified using two commercial vertical-cavity surface-emitting laser (VCSEL)-pumped amplifier modules (SLD-8$\times$5-4, Laserconn). 
The gain medium is Nd:YGAG crystal ($\phi$40~mm$\times$65~mm).
The output energy is 90~mJ at maximum. 
The beam diameter after the amplifier is adjusted to approximately 5~mm
for the subsequent frequency-conversion stages.

The second harmonic is produced in an LBO crystal ($10\times10\times20~\mathrm{mm}^3$).
A fraction of the second harmonic is split off and used as the pump light for amplification of the $\omega_2$ pulse.
The fourth harmonic is then generated using the first CLBO crystal ($10\times10\times15~\mathrm{mm}^3$).
The 212.56~nm pulse is produced by sum-frequency generation of the fundamental and fourth-harmonic pulses using the second CLBO crystal ($15\times15\times15~\mathrm{mm}^3$).
The $\omega_1$ pulse energy is 3~mJ, measured downstream of the $\omega_1$--$\omega_2$ combining mirror.

\subsection{Generation of the $\omega_2$ pulse}
A frequency-tunable Ti:sapphire laser oscillator is used as the seed source for the $\omega_2$ pulse.
An acousto-optic tunable filter (AOTF) installed inside the oscillator cavity acts as an intracavity bandpass filter; it defines the oscillation bandwidth, while its RF drive frequency controls the central wavelength~\cite{Wada:96,Yumoto_2008}.

Because the $\omega_2$ pulse has a much broader spectral bandwidth than the $\omega_1$ pulse, the bandwidth of the generated 122~nm pulse is predominantly determined by that of the $\omega_2$ pulse.
The Doppler width of muonium atoms at 300~K is approximately 80~GHz (FWHM).
Thus, the $\omega_2$ pulse should have a sufficiently broad bandwidth to allow the generated 122~nm pulse to cover the Doppler-broadened transition.
The hyperfine and fine-structure splittings of the $1S-2P$ transition are much smaller than this Doppler width and are therefore not spectrally resolved.

\begin{figure}[t]
\centering
\includegraphics[width=0.6\linewidth]{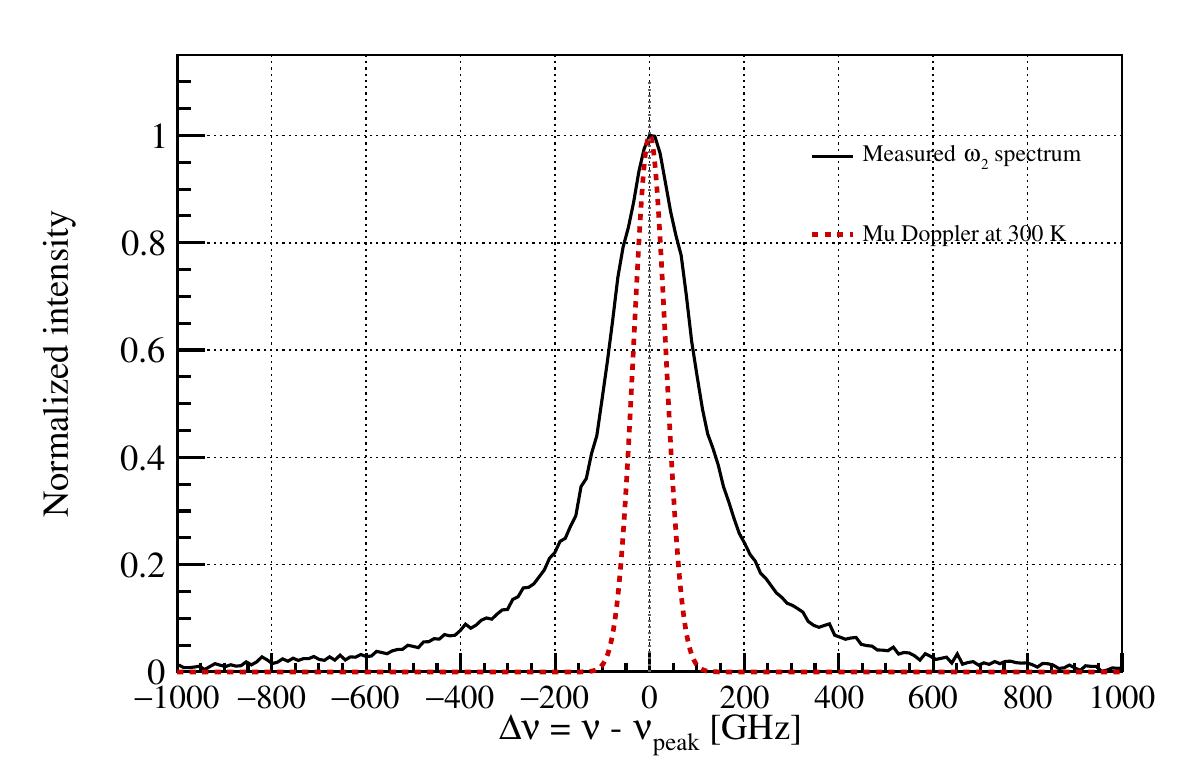}
\caption{
Spectrum of the $\omega_2$ light in the frequency domain. 
The measured wavelength spectrum is converted to frequency and shifted so that the spectral peak is located at zero. 
The black curve shows the measured $\omega_2$ spectrum, normalized to its maximum. 
The red curve shows a Gaussian profile corresponding to the Doppler-broadened thermal distribution of muonium at 300~K.}
\label{fig:spectrum}
\end{figure}

The output pulse energy of the oscillator was 100~$\mu$J.
The pulse is amplified using an OPA.
A 12~mJ portion of the 531~nm pulse from the $\omega_1$ system is used as the pump for two LBO crystals ($6\times6\times20~\mathrm{mm}^3$) in the OPA.
The $\omega_2$ pulse energy was 1.8~mJ, measured downstream of the $\omega_1$--$\omega_2$ combining mirror.
The spectral bandwidth was evaluated using a spectrometer (HR4000, Ocean Optics).
Figure~\ref{fig:spectrum} shows the measured spectrum after the OPA.
The measured FWHM was approximately 220~GHz.
The expected Doppler width of the muonium transition, converted to the corresponding $\omega_2$ frequency scale, is also superimposed.

The beam size of the $\omega_2$ pulse is adjusted using lenses upstream of the $\omega_1$--$\omega_2$ combining mirror so that the $\omega_1$ and $\omega_2$ pulses are focused at the same position by the focusing lens used for FWM.

\begin{figure*}[tbp]
\centering
\includegraphics[width=0.95\linewidth]{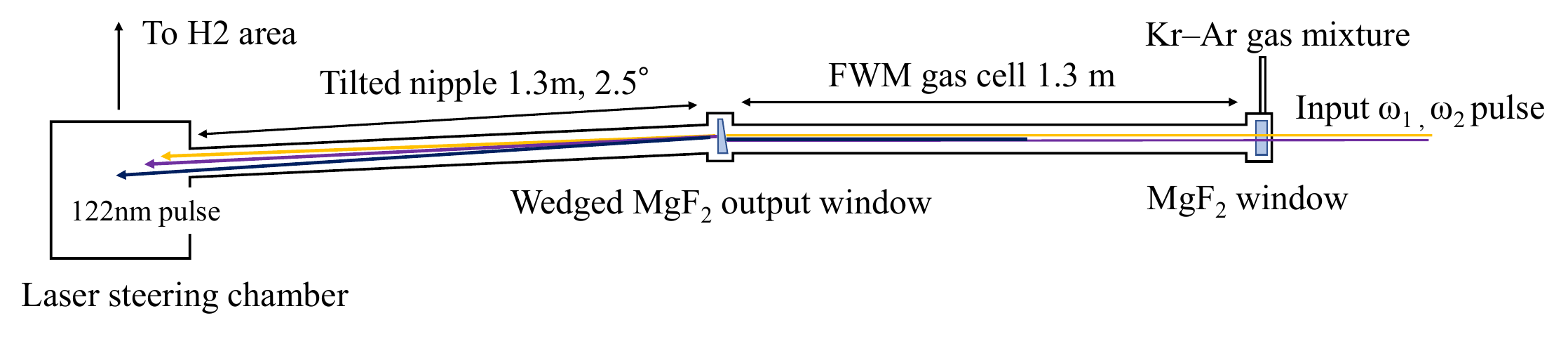}
\caption{
Schematic view of the FWM gas cell and 122~nm pulse separation geometry.
The gas cell is 1.3~m long.
The output window of the cell is a wedged MgF$_2$ window.
The gas cell is installed on the optical bench and is connected to the downstream laser steering chamber through a 1.3~m-long nipple tilted by 2.5$^\circ$ with respect to the gas-cell axis.
This geometry spatially separates the 122~nm pulse from the residual $\omega_1$ and $\omega_2$ pulses.
}\label{fig:fwm-cell}
\end{figure*}

\subsection{Coherent 122-nm pulse generation}
Figure~\ref{fig:fwm-cell} shows a schematic view of the FWM gas cell and the 122~nm pulse-separation geometry.
The $\omega_1$ and $\omega_2$ pulses are combined collinearly.
The collinear pulses are focused into the FWM gas cell using an uncoated convex lens with $f=1000~\mathrm{mm}$.
For the FWM process, the cell was filled with a Kr--Ar gas mixture at a total pressure of 4000~Pa, with a Kr:Ar pressure ratio of 1:5.5. 
The Kr:Ar pressure ratio was chosen based on the previous study~\cite{Oishi_2023}.

The gas cell length is 1.3~m.
The input window is a 3-mm-thick uncoated MgF$_2$ plate.
The output window is a wedged MgF$_2$ window with a wedge angle of 6$^\circ$ and a minimum thickness of 3~mm.
In the previous study, the output window was made of LiF~\cite{Oishi_2023}; however, MgF$_2$ was adopted in the present system because of its higher transmission at 122~nm~\cite{Oishi:2024MLFAR}.
The long cell allows the output window to be placed farther from the focal region, thereby reducing the fluence of the 122-nm pulse at the window.
The 7.2-m-long optical bench was chosen to accommodate a long FWM gas cell.
The remaining space also allows further extension of the gas cell in future upgrades.

The gas cell is connected to a laser steering chamber through a 1300-mm-long nipple tilted by 2.5$^\circ$ with respect to the gas-cell axis.
This geometry provides a sufficient propagation length for the 122~nm pulse, deflected by the wedged output window, to be spatially separated from the residual $\omega_1$ and $\omega_2$ pulses.

After the Kr--Ar gas mixture is introduced into the FWM cell, the 122~nm generation efficiency gradually decreases during operation.
This decrease is presumably related to a change in the gas mixture condition in the interaction region over time.
The gas mixture in the FWM cell is therefore automatically replaced every 3~h to maintain stable FWM efficiency.

\subsection{355~nm pulse generation}
The 355~nm pulse is generated as the third harmonic of the 1062.78~nm pulse used for the $\omega_1$ system.
The residual fundamental and second-harmonic pulses after the generation of the $\omega_1$ pulse are mixed in an LBO crystal.
The pulse energy of the generated 355~nm light is 4.5~mJ.

The temporal pulse width is measured to be 2.1~ns (FWHM) using a fast photodiode (FPD-VIS-300, Ophir) with a specified rise time of less than 0.3~ns and a 2-GHz-bandwidth oscilloscope.
The relative timing jitter of the 355 nm pulse with respect to the 12~MHz reference signal was 80--130~ps, evaluated using the same photodiode and oscilloscope. 
The 50\% crossing point of the pulse amplitude was used to define the pulse timing.
This jitter is much smaller than the 3.09~ns RF period corresponding to the 324~MHz accelerator frequency of the muon accelerator.
Together with the 122~nm excitation pulse, the 355~nm pulse contributes to the ionization-time distribution, whereas the 122~nm pulse duration could not be directly measured with the available VUV diagnostics.

A lens pair and a cylindrical lens are used to form a vertically elongated
beam spot at the muonium-production region, taking into account the
vertically elongated distribution of the emitted muonium atoms~\cite{Kamioka:2026_usm_simulation}.
The 355~nm pulse is also sent to the laser-steering chamber.

\section{VUV pulse transport and diagnostics}
\label{chap:diagnostics}

Since 122~nm pulses are strongly absorbed by air, a dedicated VUV pulse transport system under vacuum is installed in the cleanroom and connected to the muonium production chamber in the H2 area.

\begin{figure*}[t]
\centering
\includegraphics[width=0.9\linewidth]{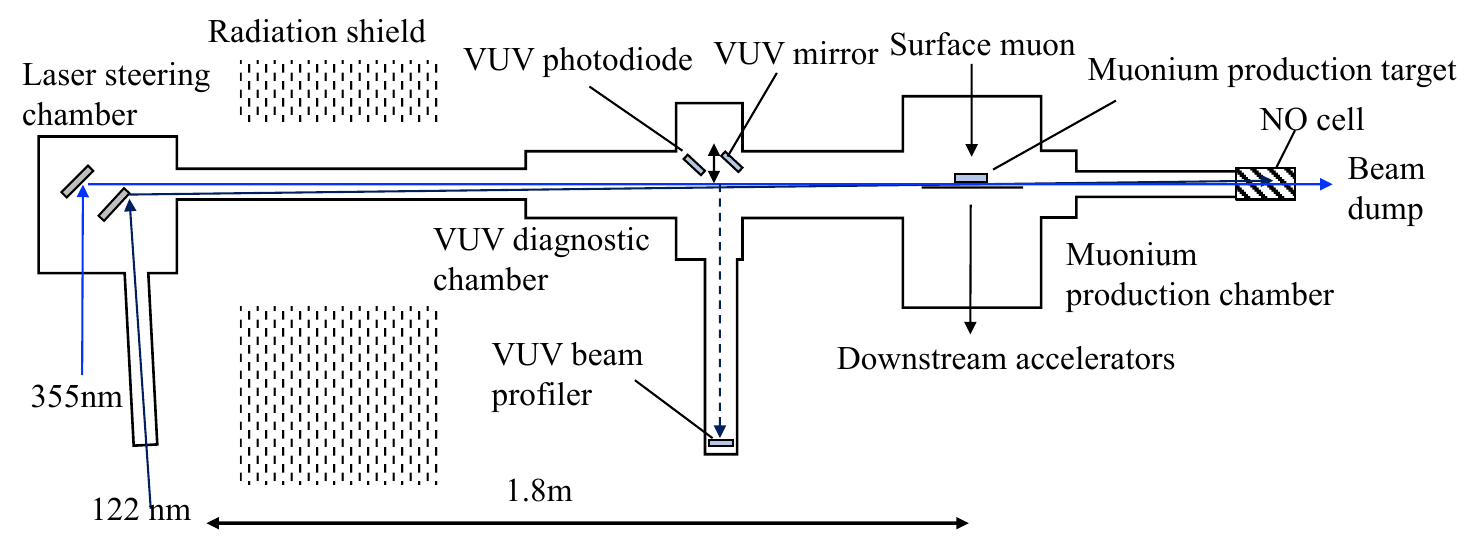}
\caption{
Schematic view of the VUV pulse transport and diagnostics under vacuum.
The 122~nm pulse produced in the FWM cell is transported to the muonium-production target region. 
The laser steering chamber contains optics for controlling the optical path. 
The VUV diagnostic chamber contains detectors for evaluating the 122~nm pulse energy and beam profile. 
The nitrogen monoxide gas cell (NO cell) is used as an online monitor of the 122~nm pulse. 
The 355~nm pulse is also transported to the muonium-production target region through the laser steering chamber and is dumped at the end of the transport line.
}
\label{fig:diagnostic}
\end{figure*}

Figure~\ref{fig:diagnostic} shows a schematic view of the 122~nm and
355~nm pulse transport and diagnostic layout, including the laser steering
chamber, VUV diagnostic chamber, muonium-production chamber, nitrogen monoxide gas cell (NO cell), and 355~nm beam dump.
The typical pressure in the system was $10^{-4}$--$10^{-5}$~Pa.
The laser steering chamber contains motorized stages and mirror mounts for remote alignment of the 122~nm and 355~nm pulses.
The VUV diagnostic chamber is equipped with detectors for characterizing
the transported 122~nm pulses.
The distance from the exit-port flange plane of the laser steering chamber to the center of the muonium-production target is approximately 1.8~m.
These components are described in more detail in the following subsections, together with the evaluation of relevant laser parameters.

\subsection{VUV pulse transport}
\label{sec:vuv_transport}
The 122~nm pulse generated in the FWM gas cell and the 355~nm pulse are transported to the muonium-production region by mirrors installed in the laser steering chamber. 
The VUV steering mirror is an enhanced-aluminum-coated concave cylindrical mirror with a radius of curvature of $R=\qty{2000}{mm}$.
Its reflectivity at 122~nm is approximately 80\%~\cite{Oishi_2023}. 
This configuration produces a vertically elongated 122~nm beam profile in the muonium target region.

Only one VUV mirror is used to minimize the transport loss of the 122~nm pulse in the present design. 
In contrast, the previous system employed two steering mirrors~\cite{Oishi_2023}, providing greater flexibility in beam alignment at the cost of additional VUV loss.
The VUV steering mirror is mounted on a vacuum-compatible motorized mirror mount (8821-L-UHV, Newport). 
To control the optical path in this single-mirror configuration, the mirror mount is fixed to a motorized translation stage (PDX1AV/M, Thorlabs). 
The closed-loop translation stage allows quantitative adjustment of the optical path along the beam axis, thereby changing the laser--target distance.

The VUV steering mirror is positioned so that the 122~nm pulse is reflected near the edge of the mirror. 
This geometry prevents the residual $\omega_1$ and $\omega_2$ pulses, spatially separated from the 122~nm pulse by the wedged output window, from being reflected toward the H2 area. 
At the same time, the reflected 122~nm pulse propagates close to the 355~nm optical path, allowing the two pulses to be transported to the muonium-production region with a small crossing angle.

\subsection{VUV diagnostics and evaluation of laser parameters}
\label{sec:vuv_energy_diagnostics}
To characterize the 122~nm pulse, the following diagnostic devices, shown in Fig.~\ref{fig:diagnostic}, are installed in the VUV diagnostics chamber.
A VUV photodiode and a VUV beam profiler, based on the designs reported in previous studies~\cite{Oishi_2023}, are used for measurements of the pulse energy and spatial profile, respectively. 
In addition, an NO cell is used as an online monitor of the 122~nm pulse during operation~\cite{MIYAKE1995265,Nakamura2014VUVTransport}.

The VUV photodiode (SXUV300C, Opto Diode) is inserted into the beam path using a vacuum linear-motion feedthrough.
The photodiode was selected for its stability under prolonged UV exposure.
A narrowband filter peaked at 122~nm (FN122-XN-2D, Teledyne Acton Optics) is placed in front of the photodiode to suppress signals from other wavelengths. 
The photodiode signal is read out with a \qty{50}{\ohm} input termination, and no reverse voltage is applied. 

The photodiode was calibrated using the 212~nm $\omega_1$ pulses. 
The calibration factor at 122~nm was derived from the responsivity ratio between 122~nm and 212~nm specified in the data sheet. 
During the June 2026 operation, the typical VUV pulse energy was \qty[uncertainty-mode = separate]{7 \pm 1}{\micro\joule}. 
The relative root-mean-square (RMS) pulse-to-pulse fluctuation was 20\%. 
The quoted uncertainty is dominated by the calibration uncertainty.

\begin{figure}[h]
\centering
\includegraphics[width=0.5\linewidth]{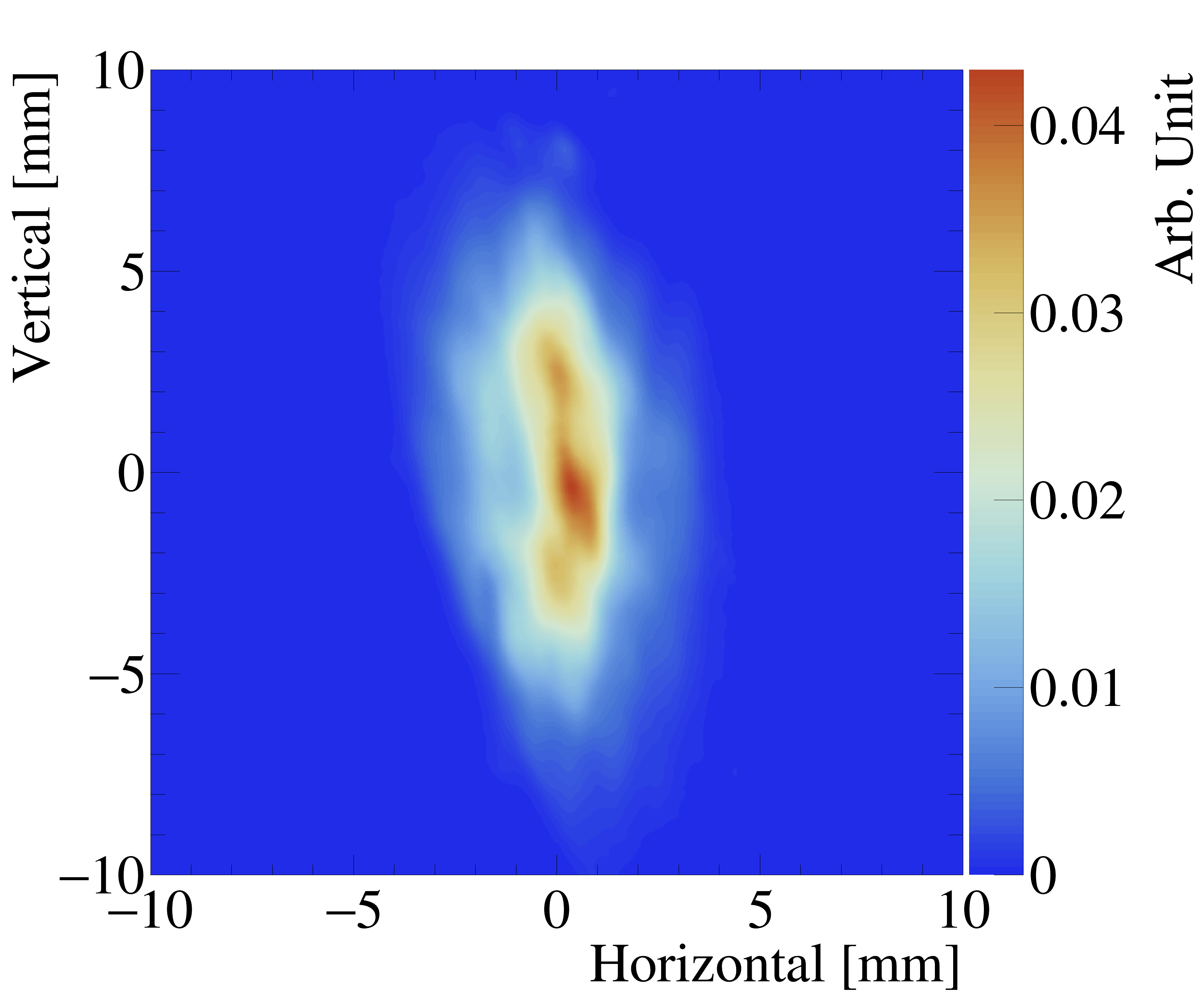}
\caption{
Transverse intensity profile of the 122~nm beam measured with the profiler, averaged over 10 shots.
The horizontal and vertical beam sizes, characterized by $D_{4\sigma}$, were $(5.6 \pm 0.2)$~mm and $13.3 ^{+0.4}_{-0.5}$~mm, respectively.
}
\label{fig:122_profile}
\end{figure}

The VUV beam profiler is based on a sodium-salicylate phosphor, which absorbs VUV light and fluoresces near \qty{410}{\nano\meter}. 
A camera records the visible fluorescence to obtain the spatial beam profile.
In the diagnostic chamber, a VUV mirror oriented at \ang{45} to the incident beam can be inserted into the beam path using a linear feedthrough to direct the beam to the profiler.
The mirror-to-profiler distance equals the mirror-to-target-center distance. 
Figure~\ref{fig:122_profile} shows a representative 122~nm laser profile. 

The beam sizes were characterized by $D_{4\sigma}$. 
The horizontal and vertical beam sizes were $(5.6 \pm 0.2)$~mm and $13.3 ^{+0.4}_{-0.5}$~mm, respectively.
The uncertainties are dominated by systematic uncertainties associated with the image-analysis procedure used to calculate $D_{4\sigma}$.
The horizontal and vertical pulse-to-pulse laser-pointing fluctuations, evaluated as the standard deviations of the beam-centroid positions, were 0.3~mm and 0.1~mm, respectively.
No pointing drift was observed during the one-hour measurement period.
The 355~nm beam profile was measured using the same procedure, and the horizontal and vertical beam sizes were $(1.9 \pm 0.1)$~mm and $(5.7 \pm 0.3)$~mm, respectively.
The spatial overlap between the 122~nm and 355~nm beams at the muonium-production region can also be evaluated using the profiler.
The measured beam profiles contribute to defining the spatial distribution of the ionized muons in the vertical and extraction directions.

\begin{figure}[htbp]
	\centering
	\includegraphics[width=0.5\linewidth]{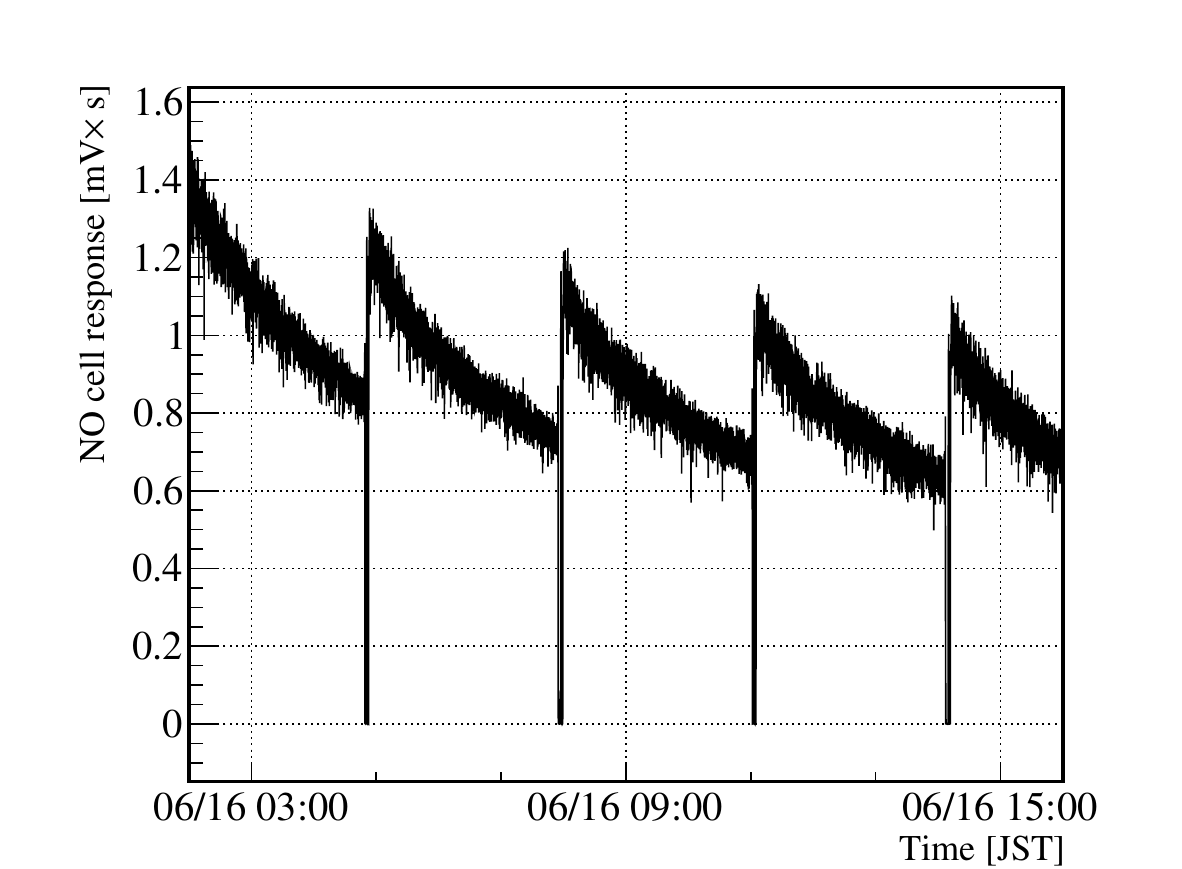}
\caption{
Time evolution of the integrated NO-cell response, recorded for each laser shot at 25 Hz during a representative 14-hour run. These data show a typical signal pattern associated with gas exchanges in the FWM cell.
}
	\label{fig:NO_cell_signal}
\end{figure}

The NO cell absorbs the 122~nm beam and yields an ionization signal for each shot. 
The structure of the NO cell, readout circuit, and the operating voltage follow the design in Ref. \cite{Nakamura2014VUVTransport}.
Its nonlinear and alignment-dependent response limits its use as a quantitative pulse-energy monitor. 
It is therefore used to monitor the VUV signal and to verify the presence of VUV pulses downstream of the muonium production region. 
Figure~\ref{fig:NO_cell_signal} shows the time evolution of the integrated NO-cell signal. 
The signal increases sharply after each FWM-gas exchange, performed at intervals of approximately \qty{3}{\hour}, and then gradually decreases until the next exchange.

\subsection{Optical path alignment and evaluation}
\label{sec:vuv_optical_path}
The laser pulses are propagated through the muonium-production region between a mesh electrode located immediately downstream of the muonium-production target and a second mesh electrode located approximately 10~mm downstream.
The spatial overlap of the laser pulses determines the ionization region and hence the initial spatial distribution of the generated ultraslow muons for the subsequent electrostatic extraction and transport.
Therefore, a well-defined and reproducible laser alignment is required.
For this purpose, the alignment was performed as follows.

A guide laser was installed and aligned to pass through the center of the muonium-production chamber and parallel to the target surface.
Before vacuum evacuation, the 355~nm pulse optical path was aligned to the guide-laser path. 
The 122~nm beam was aligned with the 355~nm beam at the profiler, thereby establishing their overlap at the target center. 
The offsets between the centroids of the two beam profiles were evaluated to be $(0.65\pm0.03)$~mm and $(0.46\pm0.04)$~mm in the horizontal and vertical directions, respectively.

During beam operation, the optical path was adjusted as needed. 
Based on the initial reference alignment, the laser--target distance is estimated to have remained between 2 and 4~mm.
Using the overlap position at the profiler and the estimated 122~nm pulse reflection point in the laser steering chamber, the crossing angle between the two beams was determined to be \qty[uncertainty-mode = separate]{0.57 \pm 0.15}{\degree}.

\section{Commissioning of the laser system}
\label{chap:beamtime}
The commissioning of the laser system was performed from June 12 to June 23, 2026.
The proton beam power at the MLF was 770~kW.

\subsection{Experimental setup}

The experimental setup was based on that shown in Ref.~\cite{wada2026diagnostic57kevmuon}.
It consisted of a muonium production target, an electrostatic immersion lens, and a diagnostic line for a low-energy muon beam, referred to as the UDL.
The surface muon beam from the H-line was stopped in a laser-ablated aerogel target, where muonium was formed~\cite{10.1093/ptep/ptu116,10.1093/ptep/ptaa145}.
The muonium atoms are emitted from the target into vacuum on a microsecond time scale~\cite{10.1093/ptep/ptz030}.
The laser was fired 1.3~$\mu$s after the arrival of the first surface-muon bunch at the target.
The ionized particles were electrostatically accelerated to 5.7~keV using an immersion lens, called the Soa lens~\cite{soalens}.
The particles were selected using an electrostatic mirror and a magnetic bend.
A single-anode microchannel plate (MCP) located at the downstream end of
the UDL was used to measure the arrival time of the particles.
An MCP-based beam profiler was used to measure the transverse profile of
the 5.7~keV muon beam~\cite{KIM201822}.
The timing of the laser pulses at the target was inferred from the signal of a photodiode located approximately 0.9~m downstream of the target, which monitored the 355~nm pulse transmitted through the NO cell.

Throughout the beam time, the 122~nm pulse energy gradually decreased even after periodic replacement of the gas mixture.
The decrease was presumably associated with degradation of optical components during VUV operation, as reported in related VUV systems~\cite{Heaney:77,Fiebrandt_2020}.
To recover the 122~nm pulse energy, the common optical axis of the $\omega_1$ and $\omega_2$ pulses was slightly shifted from time to time so that the 122~nm pulse passed through and was reflected from clean areas of the optical surfaces along its path.
After each adjustment, the spatial overlap between the 122~nm and 355~nm pulses was confirmed using the beam profiler described above.
Apart from this gradual decrease in the 122~nm pulse energy, the laser system
was routinely operated for muonium ionization measurements during the two-week
beam time.

\begin{figure*}[tbp]
\centering
\includegraphics[width=0.9\linewidth]{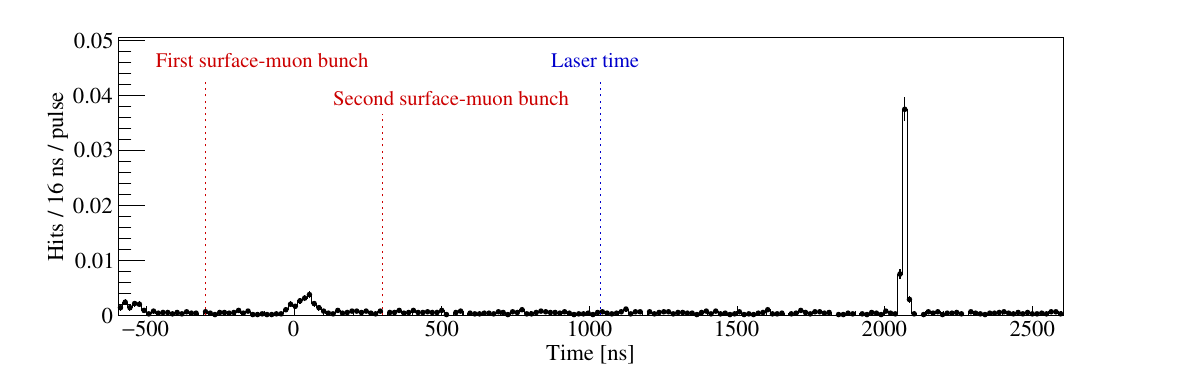}
\caption{
Timing spectrum of the MCP signal.
A pulse-height selection was applied to suppress background events.
The time origin is defined as the mean of the arrival times of the double surface-muon
bunches at the muonium production target.
The vertical lines indicate, from left to right, the arrival times of the first and second surface-muon bunches at the muonium-production target and the arrival time of the laser pulse.
A clear peak attributed to 5.7~keV muons is observed at $t=2070~\mathrm{ns}$.
Peaks observed around the time origin are attributed to signals from positrons originating from the surface muon beamline~\cite{wada2026diagnostic57kevmuon}.
}\label{fig:udl_timing_spectrum}
\end{figure*}

\subsection{Results}
Figure~\ref{fig:udl_timing_spectrum} shows the time spectrum of the detector signal.
A pulse-height selection was applied to the MCP signals to select muon-candidate events~\cite{OTANI2019162475}.
Given the double-bunch structure of the J-PARC muon beam, the time origin is defined as the mean of the arrival times of the two surface-muon bunches at the muonium production target.
The arrival time of the laser pulses at the muonium production target is indicated by the blue dashed line.
A clear peak is observed at $t=2070~\mathrm{ns}$.
The signal yield was strongly reduced when the laser frequency was detuned from the resonance condition used for normal operation.
This peak is therefore identified as the ultraslow-muon signal extracted at 5.7~keV.
During the muon beam-profile measurement, the maximum 5.7~keV muon signal rate was
$6~\mu^+/\mathrm{s}$ after the event selection, while the background rate evaluated without laser
irradiation was approximately $2\times10^{-2}~\mathrm{counts}/\mathrm{s}$.

\section{Discussion}
\label{chap:discussion}
The commissioning demonstrated that the 122~nm excitation pulse and the 355~nm ionization pulse were successfully produced and transported to the muonium production target under beamline conditions in the H2 area.
The alignment procedure for the 122~nm and 355~nm pulses to the muonium production region using the developed diagnostics was also established.
The observation of the 5.7~keV muon signal confirms the integrated operation of the ionization laser system with the low-energy muon transport line.

The observed 5.7~keV muon signal rate was
$6~\mu^+/\mathrm{s}$, whereas that reported in the previous
muon acceleration demonstration experiment was
$0.05~\mu^+/\mathrm{s}$~\cite{PhysRevLett.134.245001}.
The MCP was located downstream of a mesh electrode in the electrostatic mirror,
whose transmission is estimated to be approximately 60\% from its open-area ratio.
Taking into account this loss, which was not present in the previous measurement,
the muon rate available for injection into the downstream accelerator is estimated
to be approximately 200 times higher.

With this improved muon signal rate, beam measurements such as a Q-scan, a standard method for evaluating the transverse phase-space distribution of a beam, can be performed with a better signal-to-noise ratio, enabling a more reliable characterization of the muon beam. 
The laser parameters measured with these diagnostics will also be used to quantitatively characterize the ultraslow-muon source, which defines the initial beam distribution for the subsequent acceleration.
This information will provide input for an end-to-end simulation of the muon accelerator system.

Further optimization of the laser and surface-muon parameters will be performed to increase the ultraslow-muon rate. 
The improved rate enables more efficient optimization of the laser
parameters using the muon signal rate as an indicator.
A quantitative characterization of the muon signal rate, including its dependence on the laser parameters evaluated with these diagnostics, remains to be performed. 
This will also require a better understanding of the characteristics of the primary muon beam.
Simulations suggest that higher 122 nm and 355 nm pulse energies would enable efficient ionization over a larger muonium interaction volume~\cite{Kamioka:2026_usm_simulation}.
Upgrades of the ionization laser system are therefore planned.
These muon intensity upgrades will be important for future H2-area studies, including a demonstration of muon imaging using an accelerated muon beam~\cite{Shimomura:2025_muon_imaging}.

\section{Conclusion}
\label{chap:conclusion}
A muonium ionization laser system has been developed and installed in the H2 area of J-PARC MLF.
The laser system produces coherent 122~nm pulses to excite muonium to the 2$P$ state.
It also produces 355~nm pulses to photoionize the excited muonium.
The pulse energy of the 122~nm pulse is \qty[uncertainty-mode = separate]{7 \pm 1}{\micro\joule}. 
The pulse energy of the 355~nm pulses is 4.5~mJ.
A VUV transport system and diagnostics have also been installed in the experimental area.

The laser system was commissioned by demonstrating muonium ionization.
The muon rate available for injection into the downstream accelerator was approximately two orders of magnitude higher than that in the previous muon acceleration demonstration.
The commissioned system provides a platform for further intensity upgrades and upcoming muon accelerator studies in the H2 area.

\section*{Declaration of competing interest}

The authors declare that they have no known competing financial interests or
personal relationships that could have appeared to influence the work reported
in this paper.

\section*{Acknowledgments}
The authors would like to thank the J-PARC muon section
staff for their support in the conduct of the experiment at
J-PARC MUSE. 
The authors thank the participants of the H2-area beam time for their support
during data taking.
This work was supported in part by JSPS KAKENHI Grants No. JP22K21350,
No. JP23K13131, 
No. JP26K17183, 
No. JP26K21728
and 
JST K program No. JPMJKP24J4. 
The experiment was performed at the Materials and Life Science Experimental Facility of J-PARC under a user program 2011MS06.
We sincerely thank the Mechanical Engineering Center of KEK for supporting the design and manufacture of the experimental instruments.

\section*{Declaration of generative AI and AI-assisted technologies in the manuscript preparation process}

During the preparation of this work, the authors used ChatGPT (OpenAI) to assist with language editing of the manuscript. The authors reviewed and edited the output as needed and take full responsibility for the content of the published article.



\bibliographystyle{elsarticle-num} 
\bibliography{mybibfile}


\end{document}